\documentclass[%
 reprint,
 amsmath,amssymb,
 aps,
]{revtex4-2}

\usepackage{xcolor}

\usepackage[normalem]{ulem}

\usepackage{amsmath}
\usepackage{multirow}
\usepackage{graphicx}
\usepackage{dcolumn}
\usepackage{bm}

\begin{document}

\preprint{APS/123-QED}

\title{New Limits on Leptophilic Axionlike Particles and Majorons from ArgoNeuT}

\author{Enrico Bertuzzo}
\email{bertuzzo@if.usp.br}
\author{Ana Luisa Foguel}%
 \email{afoguel@usp.br}
 \author{Gabriel M. Salla}%
 \email{gabriel.massoni.salla@usp.br}
 \author{Renata Zukanovich Funchal}%
 \email{zukanov@if.usp.br }
\affiliation{%
 Instituto de F\'isica, Universidade de S\~ao Paulo, C.P. 66.318, 05315-970 S\~ao Paulo, Brazil
}%

\date{\today}

\begin{abstract}
Axionlike particles are among the most studied extensions of the standard model. In this Letter we study the bounds that the ArgoNeuT experiment can put on the parameter space of two specific scenarios: leptophilic axionlike particles and Majorons. We find that such bounds are currently the most constraining ones in the $(0.2 - 1.7)$ GeV mass range.
\end{abstract}

\maketitle

\textbf{\textit{Introduction.}}---Axionlike particles (ALPs) are among the best motivated additions to the standard model (SM) particle content (see~\cite{Ringwald:2014vqa} for a review). They emerge every time an exact or approximate global $U(1)$ symmetry is spontaneously broken, forcing their coupling to fermions to be derivative and to gauge bosons to involve a dual field strength. Notable examples are the axion~\cite{Peccei:1977hh,Weinberg:1977ma,Wilczek:1977pj}, familons~\cite{Davidson:1981zd,Froggatt:1978nt,Wilczek:1982rv,Jaeckel:2013uva} and the Majoron~\cite{Chikashige:1980ui,Gelmini:1980re}. They are also expected to appear in string theory. 

Besides purely theoretical motivations, ALPs can also help to explain data. Very light ALPs, with mass $m \ll 1$ eV, may be a good cold dark matter candidate~\cite{Preskill:1982cy,Abbott:1982af,Dine:1982ah,Arias:2012az}, may explain the Universe anomalous $\gamma-$ray transparency puzzle~\cite{Mirizzi:2009aj,Kohri:2017ljt} and may alleviate the 
tension between the values of the Hubble constant determined at low and 
high redshift~\cite{DEramo:2018vss}. Heavier ALPs, in the MeV-GeV mass range, may explain possible deviations in $(g-2)_e$ and $(g-2)_\mu$~\cite{Bauer:2019gfk,Cornella:2019uxs}. As there is a broad 
landscape of models, it is crucial to probe a wide range of masses and couplings in  experiments.

A crucial feature of the ALPs couplings to matter is that they are all suppressed by a factor $f$, the scale at which the associated $U(1)$ is spontaneously broken. The ALPs are thus typically long-lived and their decays can be detected far from their production point. High intensity fixed-target-produced beam experiments where a detector is located sufficiently far away from the target can thus be used to search for them. 
The ArgoNeuT~\cite{Anderson:2012vc} experiment exactly fits the profile.
The purpose of this Letter is to show that the data taken at the ArgoNeuT detector between 2009 and 2010 allow us to look for ALPs interacting with charged leptons and  place bounds in a region of parameter space previously unconstrained by other experiments.\\
\textbf{\textit{Models}}.---In this work we are going to consider two specific classes of ALPs: (i) {\it leptophilic} ALPs ($\ell$ALPs), that couple only to charged leptons and photons and (ii) the Majoron~\cite{Chikashige:1980ui,Gelmini:1980re}, associated with the $U(1)_\ell$ lepton number and responsible for the dynamical generation of right-handed neutrino masses that, in turn, generate Majorana active neutrino masses via the seesaw mechanism.

For $\ell$ALPs, we will focus on the interaction Lagrangian
\begin{equation}\label{eq:Lint_leptons}
{\cal L}_{a\ell\ell} = \frac{\partial_\mu a(x) }{2 f} \bar{\ell} \gamma^\mu \left( C_V + C_A\, \gamma_5\right)\ell +E_\gamma \frac{\alpha_{EM}}{4\pi} \frac{a(x)}{f} F \tilde{F}  \, ,
\end{equation}
where $\ell \equiv (e \; \mu \; \tau)^T$ are the charged lepton flavor fields, $C_{V, A}$ are matrices in flavor space [where both matrices can have off-diagonal entries but only $C_A$ can have diagonal entries due to the CP-odd nature of $a(x)$], $F$ and $\tilde{F}$ are the usual electromagnetic field strength and its dual (with $\tilde{F}^{\alpha\beta} = \epsilon^{\alpha\beta\mu\nu}F_{\mu\nu}/2$), while $E_\gamma$ is an ${\cal O}(1)$ number that depends on the UV completion. Additional couplings are negligible: the tree-level one to neutrinos because of the smallness of neutrinos masses, the one to quarks due to loop-suppression (we can easily use Ref.~\cite{Bauer:2020jbp} to show that it is irrelevant for our purposes). 

In the case of the Majoron (denoted by $J$), the most relevant interaction for our purposes is the loop-level coupling to charged leptons~\cite{Pilaftsis:1993af,Broncano:2002rw,Heeck:2019guh}. In the notation of Eq.~\eqref{eq:Lint_leptons} we have
\begin{equation}\label{eq:Lint_Majoron}
C_V^J = \frac{f\, K}{8\pi^2 v}, ~~~ C_A^J = -\frac{f}{8\pi^2 v} \left( K - \frac{{\rm tr}(K)}{2} \right),
\end{equation}
where $K \equiv M_D M_D^\dag/(v f)$, with $M_D$ the Dirac neutrino mass matrix and $v$ the Higgs vacuum expectation value. Additional interactions are present between the Majoron $J$ and neutrinos, photons, and quarks. The first two are irrelevant for us because they are either suppressed by $m_\nu/f\lesssim {\cal{O}}( 10^{-10})$ (where $m_\nu$ is the diagonal neutrino mass matrix~\cite{Chikashige:1980ui,Schechter:1981cv}) or generated at two-loops~\cite{Pilaftsis:1993af,Broncano:2002rw,Heeck:2019guh}. The coupling to quarks is instead generated at one-loop, as in Eq.~\eqref{eq:Lint_Majoron}, and induces interactions with mesons and nucleons~\cite{Lessa:2007up,GrillidiCortona:2015jxo}. We will see shortly that this coupling gives subdominant effects with respect to the one to charged leptons. Furthermore, we note that the matrix $K$ must satisfy $|K_{ij}| \leq {\rm tr}(K)/2$ (see Ref.~\cite{Heeck:2019guh} for more details).\\
\textbf{\textit{The ArgoNeuT experiment}}.---The ArgoNeuT detector employed the neutrino at the main injector (NuMI) beam line at Fermilab, created by 120 GeV protons, about 87\% of which interacted in the graphite target and the remaining 13\% interacted with a hadron absorber positioned 715 m downstream. The detector~\cite{Anderson:2012vc} was a 0.24 ton liquid argon time projection chamber (LArTPC) placed  100 m underground in the NuMI low energy beam line (neutrino energies in the 0.5--10 GeV range)  at 1033 m from the target and immediately in front of the MINOS Near Detector (MINOS-ND)~\cite{Anderson:2012vc,Adamson:2015dkw,MINOS:2008hdf}. The MINOS-ND was used as a spectrometer for muons exiting the ArgoNeuT detector. The active volume of the TPC was $40 \times 47\times 90$ cm$^3$ (vertical, horizontal, and beam direction) and a total of $1.25 \times 10^{20}$ protons-on-target (POT) were collected with both ArgoNeuT and  MINOS-ND operational. 
ArgoNeuT has contributed to the search for millicharged particles~\cite{ArgoNeuT:2019ckq} and provided the most stringent limit to date on tau-coupled Dirac heavy neutral leptons with a mass between 280 and 970 MeV~\cite{ArgoNeuT:2021clc}. More recently, a search for heavy QCD axions was presented~\cite{ArgoNeuT:2022mrm}. Given the similarity between our models and the QCD axion, we will base our analysis on the last search.

For what concerns us, muons generated in the decays of new physics particles were searched as a pair of minimally ionizing particles (MIPs) in the ArgoNeuT detector. The $\mu^+  \mu^-$ pair typically exits the detector and reach the downstream MINOS-ND, where the muons separate due to a magnetic field. Since the angular resolution of ArgoNeuT is about $3^\circ$, searches were conducted for three topologies: (i) muons produced inside ArgoNeuT, reconstructed as two separate tracks in both ArgoNeuT and the MINOS-ND; (ii) muons produced inside ArgoNeuT, reconstructed as a single track by ArgoNeuT and as two separate tracks by the MINOS-ND; (iii) muons produced in the 63 cm in front of the ArgoNeuT detector, reconstructed as a single track by ArgoNeuT and as two separate tracks by the MINOS-ND. The last topology takes into account that during data taking the MINERvA detector~\cite{MINERvA:2013zvz} was under construction, leaving only the quoted free space in front of the ArgoNeuT detector. 

A series of cuts (to be discussed later) were applied to identify the different topologies. The selection efficiency has been shown for events generated inside the detector or up to 50 cm upstream (see Fig. 3 of~\cite{ArgoNeuT:2022mrm}). Since no efficiency has been provided for the first 13 cm between the MINERvA and ArgoNeuT detectors, we will disregard this region in our analysis. The efficiency for the first two topologies is about 50\% for ALPs with energy above $10$ GeV. It diminishes for lower energies because the muons produced in the decay may not reach the MINOS-ND. For the events generated in the cavern upstream, the efficiency is lower, around $30\%$ (because the muons are less likely to be reconstructed as a single track) and diminishes with the distance between the muon production point and the detector. As for the backgrounds, the dominant source is charged-current muon neutrino interactions from the beam~\cite{ArgoNeuT:2022mrm}, that can mimic the double track topology (either producing two charged pions or a single muon and low energy protons). The total expected background is $0.1 \pm 0.1$ events~\cite{ArgoNeuT:2022mrm}. The experimental collaboration also quotes the following systematic uncertainties: $3.3\%$ and $0.4\%$ for the muon reconstruction inside ArgoNeuT and MINOS-ND, respectively; $1\%$ in the POT calculation; $2.2\%$ in the determination of the total volume of argon. 
Zero muon events were measured after the cuts, compatible with the expected background.\\

\vspace{-12pt}
\textbf{\textit{ALPs production and decays}}.---We now describe how the ALP models described above can generate a signal in the ArgoNeuT detector:
\begin{figure*}[tb]
\includegraphics[width=.49\textwidth]{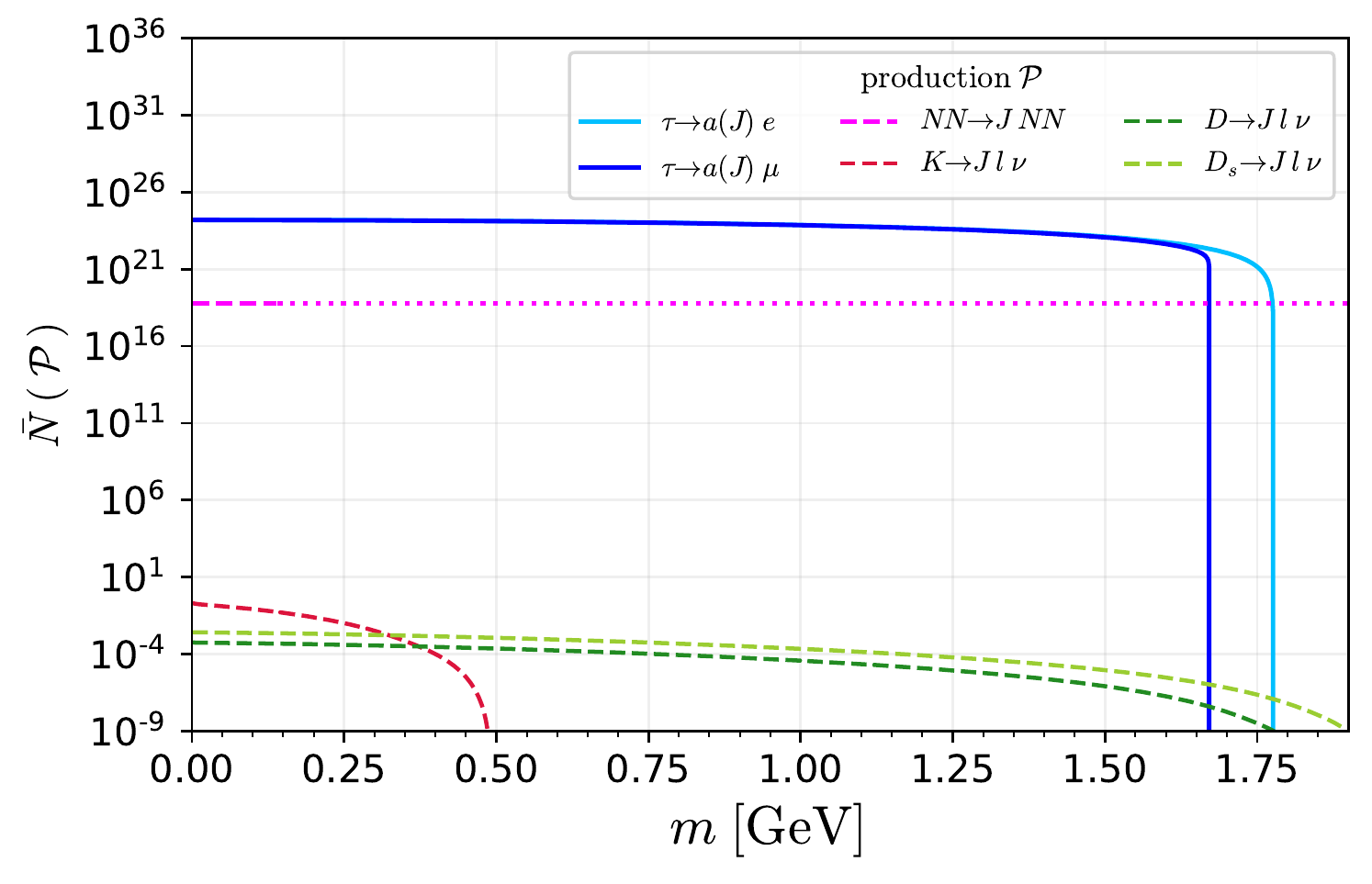}
\includegraphics[width=.49\textwidth]{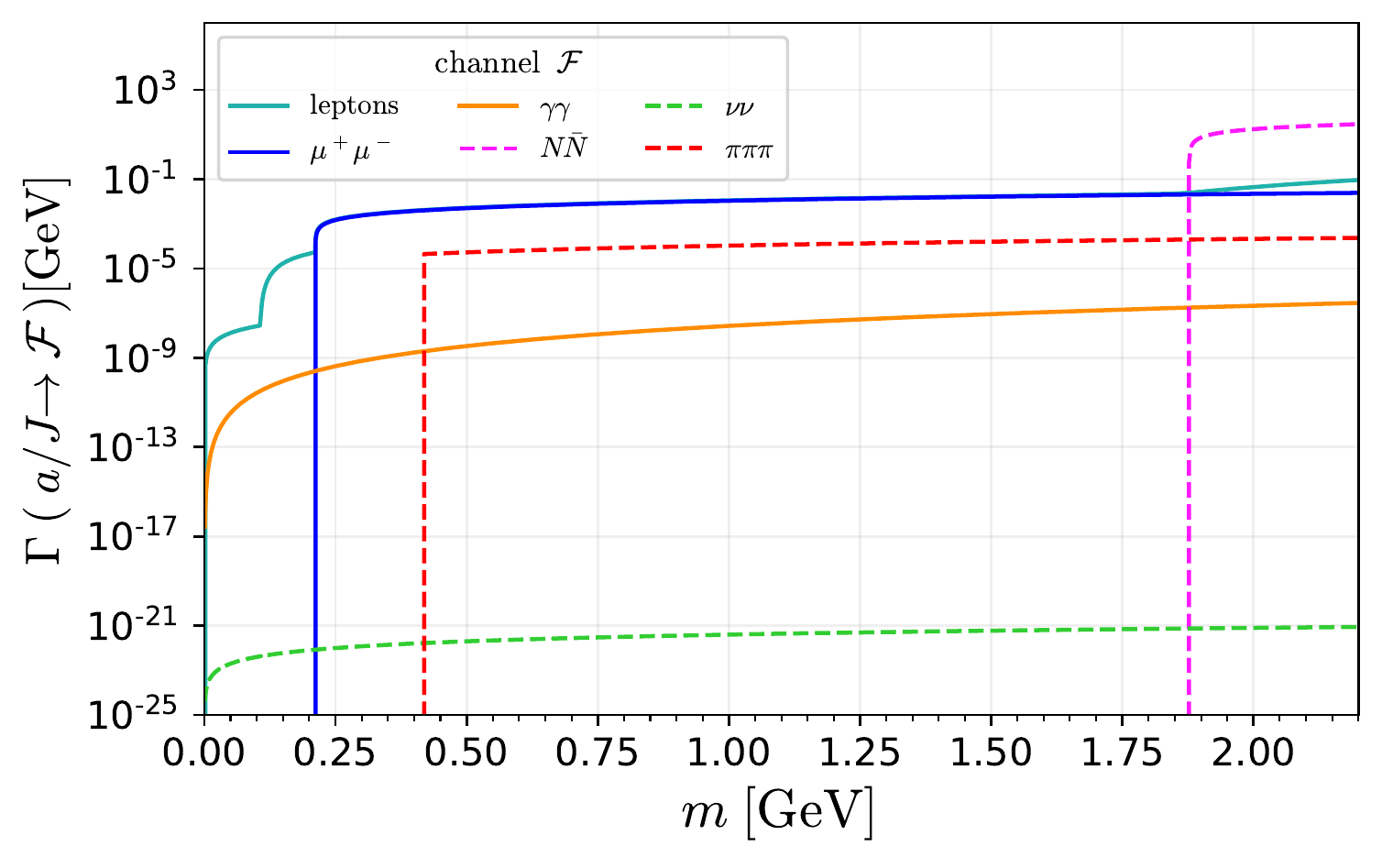}
\caption{\label{fig:prod_decay} Left panel: The solid curves show the common production mechanisms for both $\ell$ALPs and Majorons, while the dashed ones the production mechanisms which are exclusive for Majorons. The vertical axis represents the total number of ALPs produced at the NuMI target and hadron absorber for $1.25 \times 10^{20}$ POT  as a function of its mass $m$. For proton bremsstrahlung we use a dashed line for $m < m_{\pi^0}$ (the region in which the computation is valid) and a dotted line for the region $m > m_{\pi^0}$ in which we are extrapolating the computation. 
Right panel: Decay widths for $\ell$ALPs and Majorons.
The dashed lines are those channels that apply to the Majoron case only. The solid light and dark blue lines represent $\ell$ALP and Majoron decay widths, which are the same for our choice of parameters. The orange line is for $a \to \gamma  \gamma$.
Here we have fixed
$f = 1 \; \mathrm{GeV}$, $|C^{ij}_V|=C_A^{o}=1$, $E_\gamma =1$ and $R_a=5$ 
($K^o= 8\pi^2 v/\text{GeV}$ and $R_J=10$).
}
\end{figure*}
\begin{enumerate}
    \item in the case of $\ell$ALPs [Eq.~\eqref{eq:Lint_leptons}], production can proceed via (i) flavor changing decays $\ell_\alpha \to \ell_\beta  \, a$ ($\alpha$ and $\beta$ denote lepton flavors) and (ii) ALPtraum, {\it i.e.} ALP production via photon-photon fusion at the NuMI production points~\cite{Dobrich:2015jyk}. The two production mechanisms are independent: the first depends only on $C_{V, A}$,
    while the second depends only on the coefficient $E_\gamma$. 
    Since we are mainly interested in the former, we will not consider ALPtraum. Our bound will thus be conservative: the true excluded region will necessarily be larger than the one we will compute in this Letter. Since we need $m \geq 2 \, m_\mu$ to allow for the $a \to \mu^+ \mu^-$ decay, the production channels we will examine are $\tau \to \mu a$ and $\tau \to e a$, with $\tau$'s produced mainly via $D$ meson decays;
    \item in the case of the Majoron, production can proceed directly via p-nucleus interaction or indirectly via decays of short-lived particles produced at the interaction point. The production modes are: (i) flavor changing lepton decays $\ell_\alpha \to \ell_\beta  J$, (ii) meson decays $M \to \ell \nu J$, (iii) four-body lepton decays $\ell_\alpha \to \ell_\beta \nu \nu J$, (iv) resonant $pp$ scattering and (v) proton bremsstrahlung $p+{\rm target} \to J + p +{\rm target}$. The dominant production channels are $\tau \to \mu J$ and $\tau \to e J$, with $\tau$'s produced mainly via charmed meson decays. All the other production mechanisms are subdominant: the decays $M \to \ell \nu J$ and $\ell_\alpha \to \ell_\beta \nu \nu J$ both proceed via Majoron bremsstrahlung off the neutrino leg and are thus suppressed by $|m_\nu/f|^2$; resonant $pp$ scattering is not relevant because, as we will see, the interesting Majoron mass range for our bound is $m \lesssim 2$ GeV and the center-of-mass energy for $pp$ collisions generated by the NuMI beam is around 15 GeV; finally, the number of Majoron produced via proton bremsstrahlung can be estimated~\cite{CHARM:1985anb} by scaling the number of $\pi^0$ produced by the NuMI beam (about $4.5$ per POT~\cite{DeRomeri:2019kic}) by the square of the small ratio between the Majoron-nucleon and pion-nucleon couplings. Numerically, this contribution turns out to be suppressed and is expected to close for $m \gg m_{\pi^0}$.
\end{enumerate}
We show in Fig.~\ref{fig:prod_decay} (left panel) the number of $\ell$ALPs and Majorons produced at the NuMI target and hadron absorber, considering $1.25\times 10^{20}$ POT and normalizing the result to $f = 1$ GeV. The solid lines denote common production mechanisms between the two models, while the dashed lines refer to production mechanisms that are exclusive for Majorons. We have assumed degeneracy among the diagonal as well as among the off-diagonal couplings. For $\ell$ALPs this means that we take $C_A^{ii} \equiv C_A^{d}$, $C_A^{ij} \equiv C_A^{o}$, $C_V^{ij} \equiv C_V^{o}$ and define $R_a \equiv \vert C_A^d\vert/ \vert C_A^o\vert$, while for the Majoron we have $K_{ii} \equiv  K^d$, $K_{ij}\equiv K^o$ and $R_J \equiv \vert K^d\vert/ \vert K^o\vert$. We also considered that $C_A^{o}=-C_V^{o}$. Using Eq.~\eqref{eq:Lint_Majoron} we easily find a correspondence between the $\ell$ALP and Majoron couplings: $\vert C_A^o \vert= (f/8\pi^2 v) \vert K^o \vert$ and $\vert C_A^d \vert = (f/16\pi^2v) \vert K^d \vert$. We see that the choice $C_A^o = 1$ and $f=1$ GeV is translated to $K^o = 8 \pi^2 v/$GeV.\\

Once the ALPs of interest are produced, a fraction will travel in the direction of the detector and possibly produce a signature detectable in ArgoNeuT. In the case of $\ell$ALPs, the decay channels are $a \to \ell_\alpha \ell_\beta$ and $a \to \gamma \gamma$. In the case of the Majoron, possible decay channels are (i) into leptons $J \to \ell_\alpha \ell_\beta$; (ii) into nucleons $J \to N \bar{N}$; (iii) into neutrinos $J \to \nu\nu$; (iv) into three pions $J \to \pi\pi\pi$~\cite{Bauer:2017ris}. We show on the right panel of Fig.~\ref{fig:prod_decay} the decay widths of both models using the same convention for dashed and solid lines as in the left panel, with the exception of $a \to \gamma \gamma$.\\
\textbf{\textit{Signal simulation}}.---As we saw, the main production channels at the NuMI beam are $\tau$ decays, with $\tau$'s produced by $D$ mesons. We have simulated the $D^\pm$ and $D_s^\pm$ production in $pp$ collisions using \verb!PYTHIA8!~\cite{Sjostrand:2007gs}, finding $2.1\times 10^{-7}$ $\tau^+$ and $3\times 10^{-7}$ $\tau^-$ produced per POT, in agreement with~\cite{ArgoNeuT:2021clc}. The number of ALPs events inside ArgoNeuT is given by
\begin{equation}\label{eq:Nevents}
    N_{\rm evts} = \sum_i \, N_{a} f_i P_{\rm dec}^{i} ,
\end{equation}
where $i= \left\{ {\rm target}, {\rm absorber}\right\}$, $f_i= \left\{0.87, 0.13 \right\}$, $N_a$ is the number of ALPs produced and $P_{\rm dec}^{i}$ is the probability for an ALP produced at the target or absorber to give a signal in the ArgoNeuT detector. To take advantage of the MINOS-ND, we will consider as our signal only the decay $a/J \to \mu^+ \mu^-$. 

The decay probability in Eq.~\eqref{eq:Nevents} is computed as
\begin{equation}\label{eq:probability}
   P_{\rm dec}^{i} = f_{\rm geom}^i  \left( e^{- d_i/\lambda} - e^{-(d_i + l_i)/\lambda} \right)\, {\rm BR}( a/J \to \mu^+ \mu^-)\, \epsilon,
\end{equation}
where  $f_{\rm geom}^i$ is the fraction of events intersecting the detector including all cuts, $d_i$ and $l_i$ are the distances at which the ALP enters and exits the detector (computed with respect to the position of the production point), $\lambda$ is the ALP decay length $\lambda = c \beta\gamma \tau$ (with $\gamma$ the boost factor, $c\beta$ the velocity and $\tau$ the lifetime), ${\rm BR}(a/J \to \mu^+ \mu^-)$ is the branching ratio into muons and $\epsilon$ is the detection efficiency of muon reconstruction. 

For our Monte Carlo simulation, we created a \verb!FeynRules!~\cite{Alloul:2013bka} file and we fully implemented the ArgoNeuT geometry in \verb!MadDump!~\cite{Buonocore:2018xjk} (with fiducial volume $1\leq x \leq 46$ cm, $-19 \leq y \leq 19$ cm, and $z \geq 3$ cm for the drift, vertical, and beam directions, respectively). We use \verb!MadDump! for the simulation of ALP production, decays, and the calculation of $P_{\rm dec}^i$. From the simulation, we find that, for ALPs masses in the $0.2 - 1.7$ GeV range, the muons produced in the decays have average angle with respect to the beam of $\langle\theta_\mu\rangle \simeq 0.5^\circ -  2.5^\circ$, average opening angle $\langle\theta_{\mu\mu}\rangle\simeq 1^\circ - 5^\circ$, and average energy $\langle E_\mu \rangle \simeq 16-21$ GeV, in perfect agreement with what has been found by the experimental collaboration~\cite{ArgoNeuT:2022mrm}.

For the analysis, we follow as much as possible Refs.~\cite{ArgoNeuT:2021clc,ArgoNeuT:2022mrm}. For the range of ALP masses considered, we find a geometric acceptance of approximately $(8-12)\times 10^{-4}$\% and $(6-18)\times 10^{-3}$\% for events produced at the interaction point and hadron absorber, respectively. 
The double track topology is defined requiring the opening angle between muons to satisfy $3^\circ < \theta_{\mu\mu} \leq 15^\circ$, while for the single track we require $\theta_{\mu\mu} \leq 3^\circ$ and $\theta_{\mu} \leq 10^\circ$. We find that $4\%-89\%$ ($96\%-11\%$) of the events fall in the double (single)-track category for an ALP mass in the $(2 m_\mu - m_\tau)$ interval. Almost no events are lost imposing the additional cuts $\theta_{\mu\mu} \leq 15^\circ$ and $\theta_{\mu} \leq 10^\circ$ (see Appendix 1). In both cases, the muon tracks are required to be longer than $5$ cm. Since the position of the ALP decay is unknown in our simulation, we implement this requirement intersecting the trajectory of each simulated muon with the detector, keeping only those events for which the distance between the exit and entry points is larger than $5$ cm. This also guarantees that the muon pair can reach the MINOS-ND, given its larger dimensions. We find that all the simulated events produce tracks longer than 5 cm. To implement the detection efficiency given by the collaboration, we first require the ALP energy to be $E > 10$ GeV (since below this value the efficiency dramatically drops). We find that about $30$\% of the simulated events pass this cut. We then proceed in different way according to the type of events: for double and single track events which start inside ArgoNeuT we conservatively take $\epsilon = 0.4$ in Eq.~\eqref{eq:probability}; for single track events which start in the region 50 cm upstream from ArgoNeuT we conservatively take $\epsilon = 0.2$.  We assume that this efficiency encodes all detector effects that have not been applied in our simulation. Since we apply the same cuts as in~\cite{ArgoNeuT:2022mrm}, we expect the same background of $0.1\pm 0.1$ events. We present in the Appendix 2 our estimate of the systematic uncertainties. \\
\textbf{\textit{Results}}.---
\begin{figure}[tb]
\includegraphics[width=.45\textwidth]{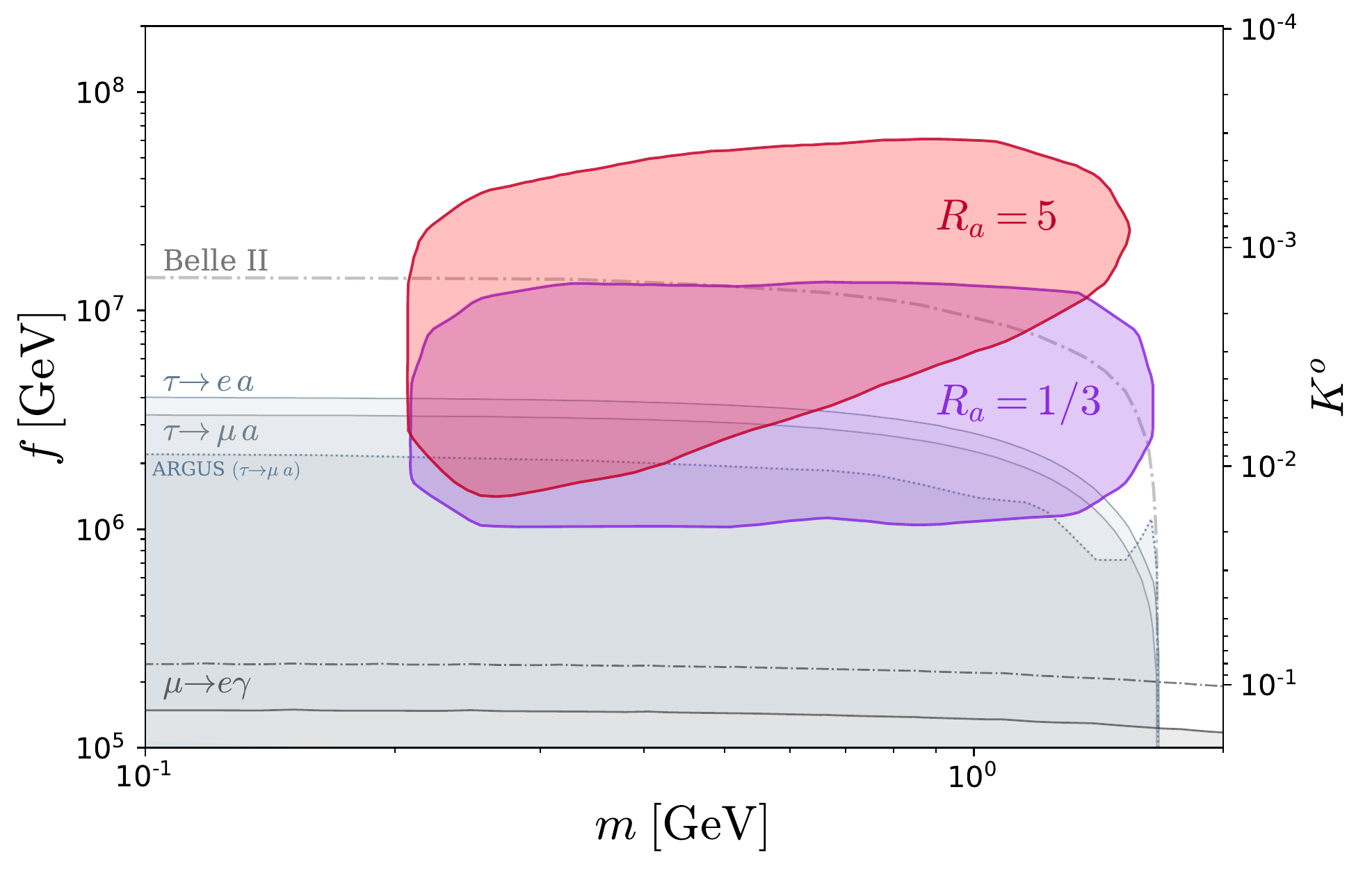}
\caption{\label{fig:money_plot} Region excluded by the ArgoNeut data for the $\ell$ALP (Majoron) model with $R_a=5$ ($R_J=10$) in red and  with $R_a=1/3$ ($R_J=2/3$) in purple at 95\% C.L. We also show in gray the regions excluded by the ARGUS experiment~\cite{ARGUS:1995bjh}, the bounds computed in \cite{Bryman:2021ilc}, the bound coming from $\mu \to e \gamma$ \cite{ParticleDataGroup:2020ssz} as well as the expected sensitive reach of Belle II (dash-dotted line above)~\cite{Yoshinobu:2017jti} and of MEG II (dash-dotted line below) \cite{MEGII:2018kmf}. Here we have fixed the other parameters as in Fig.~\ref{fig:prod_decay}.}
\end{figure}
Our main results are presented in Fig.~\ref{fig:money_plot}, where we show the 95\% C.L. exclusion due to ArgoNeuT on the $(m, f)$ plane for two choices of parameter: $R_a = 5$ and $1/3$ (corresponding to $R_J = 10$ and $2/3$, respectively \footnote{In the minimal Majoron model $R_J\geq 2/3$ follows from the inequality $| K_{ij} |< {\rm tr}(K)/2$~\cite{Heeck:2019guh}, given our choice of diagonal and off-diagonal elements. Using the Casas-Ibarra parametrization \cite{Casas:2001sr} we find that $R_J\sim 1$ would be preferred in the minimal setup, while higher values can be achieved if we allow more sterile states. To consider both situations, we choose $R_J = 10$ as a second benchmark.}). We have fixed the other parameters to $\vert C_V^{ij} \vert = C_A^o = 1$ and $E_\gamma = 1$. For this choice of parameters, we can translate between the $\ell$ALP and Majoron case using the relation $K^o = (8\pi^2 v)/f$, so that the bound can be interpreted either in terms of $f$ or $K^o$. The two possibilities are shown in the left and right vertical axis, respectively. The limit is computed following the approach outlined in  Appendix 3. The bound becomes ineffective below $m \simeq 200$ MeV (because the $a/J \to \mu^+ \mu^-$ channel closes) and above $m \simeq m_\tau$ (because the production channel closes). We also lose sensitivity for large values of $f$ (small values of $K^o$) because in this region the number of ALPs produced diminished drastically, and for small values of $f$ (large values of $K^o$) because the lifetime decreases and the $a/J$ typically decays before reaching the detector. Overall, despite its small dimensions, ArgoNeuT is able to put relevant bounds on the parameter space because $a/J$ can be abundantly produced (thanks to the large number of $D$ and $D_s$ mesons present in the beam) and the decay width is sufficiently suppressed to allow it to travel the long distance to the detector.

It is interesting to compare the ArgoNeuT bounds with other limits that constrain the same region in parameter space. Those relevant for us are the bounds on $\tau \to e +a$ and $\tau \to \mu + a$ coming from the ARGUS Collaboration~\cite{ARGUS:1995bjh}, the more recent bounds computed in~\cite{Bryman:2021ilc} (gray regions) and the bound coming from $\mu \to e \gamma$ \cite{ParticleDataGroup:2020ssz}. Future experiments will also be able to partially probe the same parameter space. In particular, the limits on the off-diagonal ALP-$e$-$\tau$ and ALP-$\mu$-$\tau$ couplings can be improved by Belle-II~\cite{Yoshinobu:2017jti}. A preliminary forecast obtained considering a luminosity of 1~ab$^{-1}$ is shown with a dashed-dotted gray line. In addition, also the MEG II experiment~\cite{MEGII:2018kmf} will be able to improve the limits on $\mu \to e \gamma$.\\
\textbf{\textit{Conclusions}}.--- We have studied how data collected at the ArgoNeuT detector can exclude the parameter space available for $\ell$ALPs and Majorons. Our main result is shown in Fig.~\ref{fig:money_plot} for the choice of parameters specified in the caption. The limits obtained in the $2 m_\mu \lesssim m \lesssim m_\tau$ mass range are the most stringent up to date and will only be partially probed by Belle II at 1 ab$^{-1}$. 

Our study can be extended to predict the reach for ALPs of the future experiments in connection to the Short Baseline Neutrino Program at Fermilab~\cite{Machado:2019oxb}. These will consists of three LAr-TPC (MicroBooNE, SBND and ICARUS), placed at 100-600 m from the interaction point. These distances are similar to the one between the hadron absorber and ArgoNeuT, so we expect they will be able, in the future, to put limits competitive with those computed in this work. We will study this point in a future publication.\\

\vspace{-0.8cm}
\begin{acknowledgments}
We thank Luighi P. S. Leal for providing us with the $\mu \to e \gamma$ limits. We also thank Luca Lista and Lucas M. D. Ramos for useful discussion.
E. B. and R. Z. F. were partially supported by Fundação de Amparo à Pesquisa do Estado de São Paulo (FAPESP) under Contract No. 2019/04837-9 and 2022/08770-9, and Conselho Nacional de Desenvolvimento Científico e Tecnológico (CNPq).
A.~L. F. and G. M. S. were fully financially supported by FAPESP under Contracts No. 2022/04263-5 and No. 2020/14713-2, respectively. R. Z. F. would like to thank Departament de Física Quàntica i Astrofísica and Institut de Ciencies del Cosmos, Universitat de Barcelona for the hospitality during the finalization of this work. G. M. S. would like to thank the Theory Group of the Deutsches Elektronen-Synchrotron (DESY) for the hospitality during the final stages of this work.

\end{acknowledgments}

\vspace{1em}

\textbf{\textit{Appendix A:  Signal efficiencies and kinematics distributions. ---}}
Here we give additional information on the impact of the various cuts on the expected signal. In Table~\ref{tab:Cuts} we show for a few benchmark masses, independently of the model, the percentage of events that survive the geometrical acceptance of the detector,
the ALP minimum energy requirement, the maximum opening angle for the muon pair and the muons maximum deviation 
from the beam direction. These cuts have been applied to all events.
The geometry acceptance is brutal, keeping only about $10^{-5}$ ($10^{-4}$)
of the signal events produced in the target (absorber).
We also see that the cuts on the angles are very weak and eliminate very few events, while around $70\%-80\%$ of the events are discarded by imposing $E>10$ GeV. 
Regarding the track length, we observed that none of the muons produce tracks smaller than $5$ cm in length, so we did not reject any event with this cut.
Note that an additional overall detector reconstruction efficiency $\epsilon$  of 40\% (20\%) is further applied to double and single track events that start inside ArgoNeuT (single track events that start upstream from the detector). We believe this conservatively takes into account all the detector effects non-included in our simulation.

\begin{table}[ht]
\def\arraystretch{1.8}
    \centering
    \vspace{4mm}
\begin{tabular}{lrrr}
\hline
           & $m_1 $ & $\quad \quad \quad m_2$ & $\quad \quad \quad m_3 $ \\ \hline
  $f_{\rm geom}^{\rm target}$ & $8.2 \times 10^{-4}$\% & $8.2 \times 10^{-4}$\% &  $12 \times 10^{-4}$\% \\
  $f_{\rm geom}^{\rm absorber}$ & $6.2 \times 10^{-3}$\% & $8.6 \times 10^{-3}$\% &  $18 \times 10^{-3}$\%  \\
  $E> 10$ GeV & 23\%& 22\% & 34\%\\
  $\theta_{\mu\mu} \leq 15^\circ$ & 99.91\%& 99.9\%& 94\%  \\
  $\theta_\mu \leq 10^\circ$  &$>99.99$\% & 100\%& 100\%\\ \hline 
\end{tabular}
        \caption{Percentage of signal events that survive the geometrical acceptance $f_{\rm geom}$ for events produced in the target or absorber and the overall kinematic cuts imposed on
        the ALP energy $E$ and muon angles $\theta_{\mu\mu}$ and $\theta_\mu$ for the three different benchmark masses $m_1 = 0.3$, $m_2 = 0.8$ and $m_3 = 1.6 \, {\rm GeV}$. }
    \label{tab:Cuts}
\end{table}

\begin{figure}[ht]
    \centering
    \includegraphics[width=.42\textwidth]{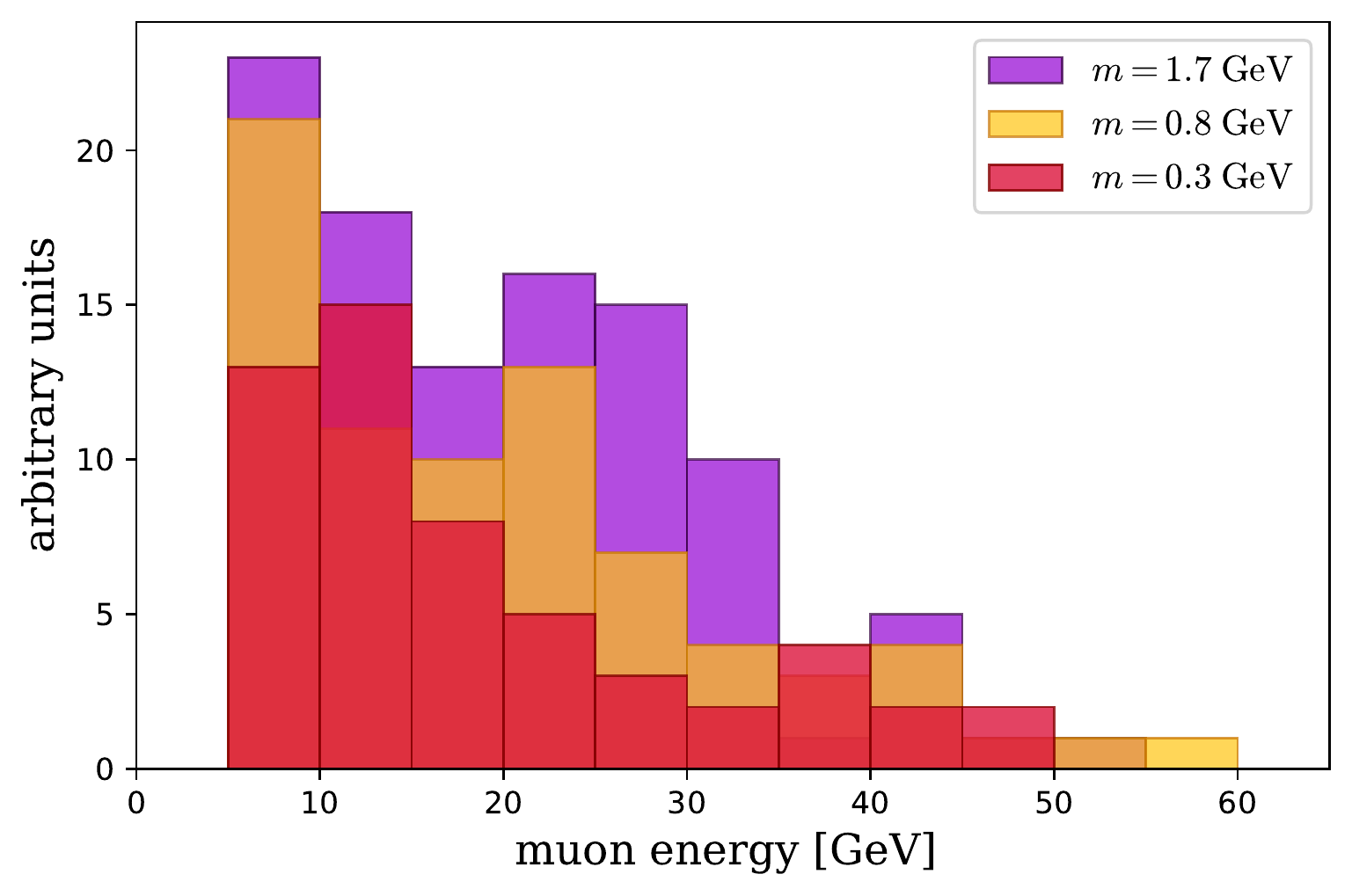}
    \includegraphics[width=.42\textwidth]{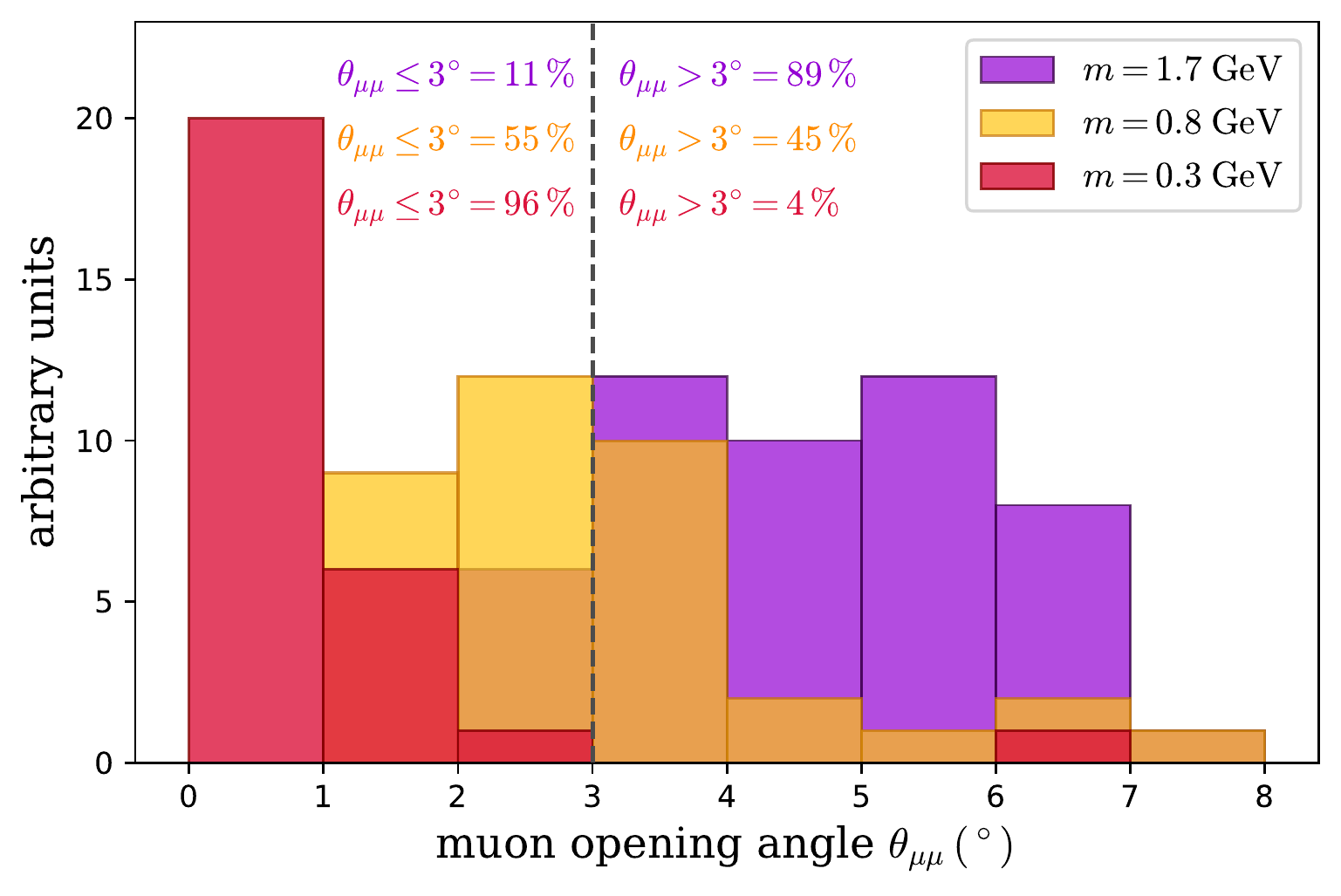}
    \includegraphics[width=.42\textwidth]{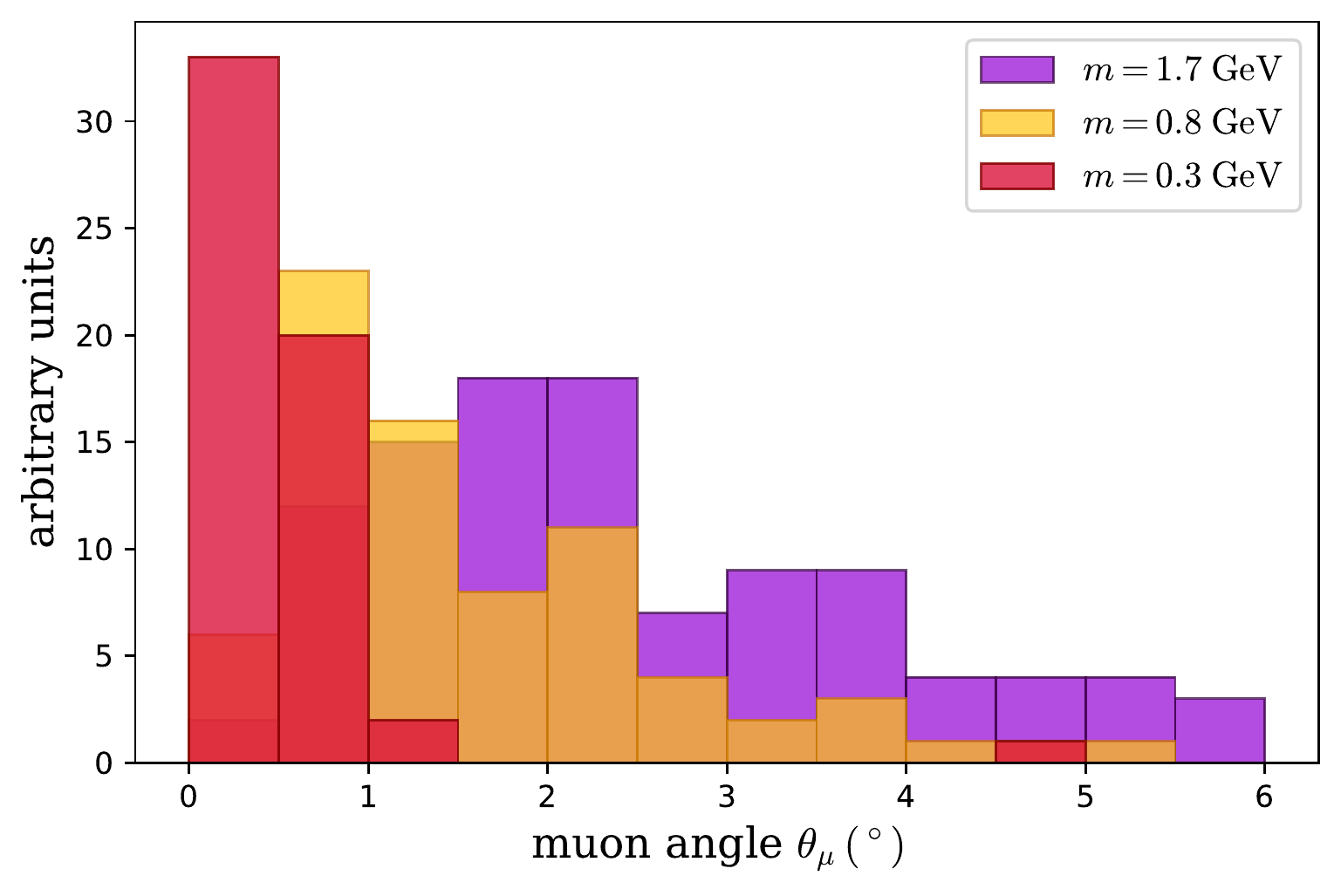}
    \caption{Distributions obtained by our simulation: muon energy (upper panel), muon opening angle (middle panel) and muon angle with respect to the beam (lower panel). The dependence on the ALP mass is highlighted showing three different benchmarks. In the middle panel we also summarize, for the three different masses, the percentage of events that fall into the single and double track signatures. See text for the cuts applied on these  simulated events.}
    \label{fig:muon_energy}
\end{figure}
\begin{figure}[ht]
    \centering
    \includegraphics[width=.42\textwidth]{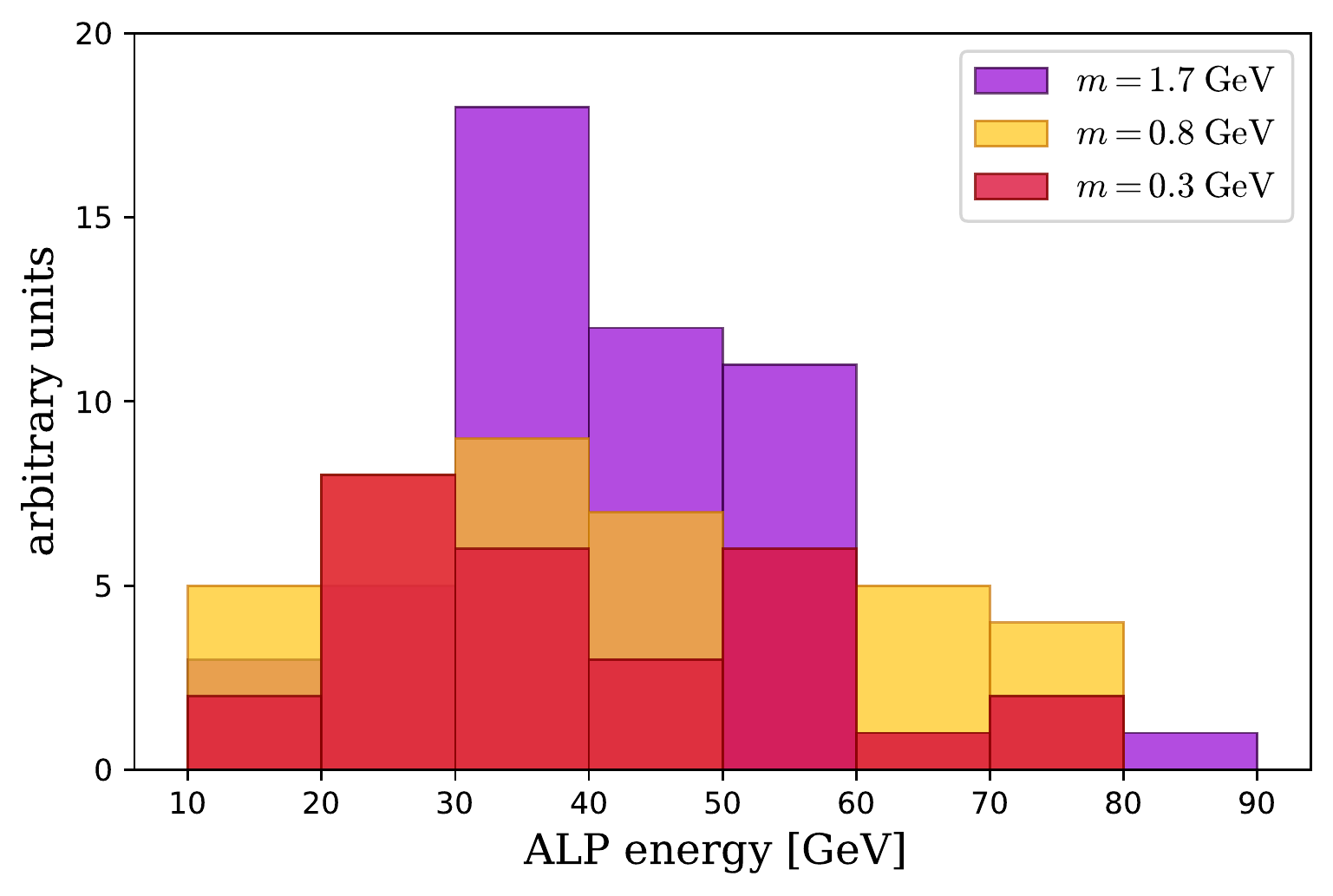}
    \caption{ALP energy distribution for three benchmark masses.}
    \label{fig:alp}
\end{figure}

\begin{table}[ht]
\def\arraystretch{1.5}
    \centering
    \vspace{4mm}
    \begin{tabular}{llcccc}
    \hline
    \multicolumn{6}{l}{$R_J=2/3$ ($R_a = 1/3$)}                       \\ \hline 
                    &   & $N_{\rm ID}$ & $N_{\rm IS}$ & $N_{\rm US}$ & $N_{\rm tot}$ \\ \cline{2-6} 
       \multirow{3}{*}{$K^o = 10^{-2}$} & $m_1 \quad$ &  $\, 0\, $  &  $\, 72.3 \, $  &  $ \, 34.7 \, $  &  $ \, 107 \, $    \\
          \multirow{3}{*}{$(f[{\rm GeV}] = 1.9 \times  10^{6})\; \; $}          & $m_2$ &  $1.0$  &  $37.9$  &  $18.2$  &  $57.1$ \\
                    & $m_3$ &    $1.6$  &  $0.5$  &  $0.3$  &  $2.4$      \\ \cline{2-6} 
   \multirow{3}{*}{$K^o = 6 \times 10^{-3} \; \;$} & $m_1$ &   $0$  &  $99.8$  &  $47.7$  &  $147.5$     \\
         \multirow{3}{*}{$(f[{\rm GeV}] = 3.2 \times 10^6) \; \;$}   & $m_2$ &    $11.4$  &  $45.9$  &  $21.9$  &  $79.2$     \\
                    & $m_3$ &   $5.6$  &  $0.9$  &  $0.4$  &  $6.9$     \\
                    \hline
    \end{tabular}
        \caption{ Number of expected
        signal events for the three different topologies: 
        $N_{\rm ID}$ (muons start inside ArgoNeuT/double track),
        $N_{\rm IS}$ (muons start inside ArgoNeuT/single track),  and $N_{\rm US}$ (muons start upstream from ArgoNeuT/single track) for fixed $R_J= 2/3$ ($R_a = 1/3$) and considering three benchmark masses $m_1 = 0.3 \, {\rm GeV}$, $m_2 = 0.8 \, {\rm GeV}$, $m_3 = 1.6 \, {\rm GeV}$ and two off-diagonal couplings $K^o$ ($f$ values). 
        We also show the total number $N_{\rm tot}$ of signal events. For all cases no events were experimentally observed. }
    \label{tab:NevtsR23}
\end{table}
\begin{table}[ht]
\def\arraystretch{1.5}
    \centering
    \vspace{4mm}
    \begin{tabular}{llcccc}
    \hline
    \multicolumn{6}{l}{$R_J=10$  ($R_a = 5$)}                       \\ \hline 
                    &   & $N_{\rm IS}$ & $N_{\rm ID}$ & $N_{\rm US}$ & $N_{\rm tot}$ \\ \cline{2-6} 
    \multirow{3}{*}{$K^o = 3 \times 10^{-3} \; \;$} & $m_1 \quad$ &  $\, 244.9 \,$  &  $\, 15.7\, $  &  $\, 117.1 \,$  &  $\, 377.7 \,$    \\
   \multirow{3}{*}{$(f[{\rm GeV}] = 6.5 \times  10^{6})\; \; $}                 & $m_2$ &  $17.3$  &  $0.6$  &  $8.3$  &  $48$       \\
                    & $m_3$ &    $0$  &  $0$  &  $0$  &  $0$      \\ \cline{2-6} 
    \multirow{3}{*}{$K^o = 10^{-3} $} & $m_1$ &   $13.8$  &  $6.7$  &  $6.6$  &  $27.1$     \\
       \multirow{3}{*}{$(f[{\rm GeV}] = 1.9 \times  10^{7})\; \; $}              & $m_2$ &    $20.6$  &  $17.7$  &  $9.8$  &  $48.1$     \\
                    & $m_3$ &   $0.2$  &  $0.9$  &  $0.09$  &  $1.2$     \\
                    \hline
    \end{tabular}
        \caption{ Same as Table~\ref{tab:NevtsR23}, but for $R_J= 10$  ($R_a = 5$). Again, we remark that no events were experimentally observed by the collaboration.}
    \label{tab:NevtsR10}
\end{table}
We also present in this Appendix some interesting distributions derived from our Monte Carlo simulation.  In Fig.~\ref{fig:muon_energy} we show the distributions of the muon energy (upper panel), the opening angle between the muon pair $\theta_{\mu\mu}$ (middle panel) and the angle between muons and beam $\theta_\mu$ (lower panel). Different colors refer to different benchmark masses. All events shown have passed  the detector geometric acceptance as well as the cuts: $E> 10$ GeV, $\theta_{\mu \mu} \leq 15^\circ$ and $\theta_\mu \leq 10^\circ$.
The detector reconstruction efficiencies have not been considered in these distributions yet. In the middle panel we also show the fraction of events that fall in the single and double track category for the three benchmark masses. In Fig.~\ref{fig:alp} we show the energy distribution of the ALPs, again for three benchmark values of the mass, obtaining an average of 38 GeV. 
  
Finally, to illustrate, Tables~\ref{tab:NevtsR23} and~\ref{tab:NevtsR10} show the final number of signal events in each of the three decay topologies we considered in the simulation, for fixed benchmark choices of $R_J$ ($R_a$), the ALP mass $m$ and the off-diagonal coupling $K^o$ ($f$ value). For lower ALP masses, the majority of the events come from the muons starting inside the ArgoNeuT detector with a single-track signature ($N_{\rm IS}$), followed by the muons that start upstream  ($N_{\rm US}$). Conversely, when the ALP masses are higher the decays inside the detector with the double track signature ($N_{\rm ID}$) dominate.

\textbf{\textit{Appendix B: Systematic uncertainties}}. --- The systematic uncertainties in our analysis were 
estimated  by changing the cuts described in the text within $1\sigma$ of the respective assumed resolution. 
We have considered an energy resolution of $6\%$~\cite{MINOS:2008hdf} and an angular resolution of $3\%$. The main systematic uncertainty comes from the Monte Carlo simulation, which we estimate to be around $21\%$, in agreement with what was found in~\cite{ACCMOR:1988pxc, E769:1996jqf,Lourenco:2006vw} and by the ArgoNeuT collaboration. We also included the uncertainties related with the effect of neutrino-induced through-going muons events in the selection efficiency~\cite{ArgoNeuT:2012jmq,Anderson:2012vc} as well as the systematics associated with the size of the ArgoNeuT detector volume~\cite{Spitz} and the number of collected POT~\cite{ArgoNeuT:2014rlj}. The impact of the systematic uncertainties on the total number of events is summarized in Table~\ref{tab:systematics}.\\
\begin{table}[hbt]
\vglue -0.8cm
    \centering
    \vspace{4mm}
    \begin{tabular}{c|c}
    \hline
         Systematic Uncertainty & Impact (\%)  \\ \hline
         Monte Carlo simulation of the flux & $21$\\
         Selection efficiency & $3.3$ \\ 
         ArgoNeuT volume  & $2.2$ \\
         POT computation & $1.0$ \\
         Energy cut & $2$\\
         Single/Double Track criteria & $1.1$ \\
         Opening angle cut & $0.4$ \\
         Maximum angle cut & $<0.001$\\
         Length of the track & $<0.001$\\ \hline
         Total & 21.5\\
         \hline
    \end{tabular}
      \caption{Impact of the systematic uncertainties on the sensitivity.}
    \label{tab:systematics}
\end{table}

\textbf{\textit{Appendix C: Statistical treatment}}. --- To determine the region excluded by ArgoNeuT, we use the following test statistics \cite{Cowan:2010js,Coloma:2022hlv}:
\begin{align}\begin{aligned} \label{eq:test_statistics}
\Lambda = {\rm min}_{\xi_S,\xi_B} \Bigg\{ 2( N_\text{pred} & - N_\text{obs})  + 2 N_\text{obs}\log \frac{N_\text{obs}}{N_\text{pred}}  + \\
& + \left(\frac{\xi_S}{\sigma_S}\right)^2 + \left(\frac{\xi_B}{\sigma_B}\right)^2 \Bigg\},
\end{aligned}\tag{C1}\end{align}
where $N_\text{obs} = 0 $ is the number of observed events by the experiment and
\begin{equation}
N_\text{pred} = (1+\xi_S) N_\text{S} + (1+\xi_B)N_\text{B},\tag{C2}
\end{equation}
with $N_\text{S}$ the simulated number of events as a function of the model parameters and $N_\text{B} = 0.1$ is the expected background computed by the collaboration. Here, $\xi_{S,B}$ are nuisance parameters associated with the simulated signal and background, and $\sigma_{S,B}$ the corresponding relative uncertainties, i.e., $\sigma_S=0.215$ and $\sigma_B=1$. The quantity $\Lambda$ in Eq. \eqref{eq:test_statistics} represents the log-likelihood ratio between the ALP and the SM model hypothesis, assuming that both are Poisson distributed. Under the assumption that $\Lambda$ is distributed as a two-dimensional $\chi^2$, we can obtain the region excluded at 95\% C.L. by demanding that $\Lambda > 3.84$~\cite{ParticleDataGroup:2020ssz}. \\

\bibliography{Leptophilic_ALP_argoneut}

\end{document}